\documentclass[journal, twocolumn]{IEEEtran}
\usepackage{stmaryrd}
\usepackage{amsfonts}
\usepackage{amsmath}
\usepackage{amssymb}
\usepackage{cite}
\usepackage{graphicx}
\usepackage{subfigure}
\usepackage{amsthm}
\usepackage{algorithmic}
\usepackage{algorithm}
\usepackage{stfloats}
\newtheorem{Theorem}{Theorem}
\newtheorem{Definition}{Definition}
\newtheorem{Lemma}[Theorem]{Lemma}
\newtheorem{Corollary}[Theorem]{Corollary}
\newtheorem{Example}{Example}

\newtheorem{Construction}{Construction}
\def\QED{\IEEEQED\vspace{0.1in}}
\newcommand{\nchoosek}[2]{\binom{#1}{#2}}
\begin{document}
%

\title{Efficiently Extracting Randomness from Imperfect Stochastic Processes}

\author{Hongchao~Zhou,
        and~Jehoshua~Bruck,~\IEEEmembership{Fellow,~IEEE}
\thanks{Hongchao~Zhou and Jehoshua~Bruck are with the Department
of Electrical Engineering, California Institute of Technology, Pasadena,
CA 91125, USA, e-mail: hzhou@caltech.edu; bruck@caltech.edu.}
\thanks{This work was supported in part by the NSF Expeditions in Computing
Program under grant CCF-0832824.}
}


\maketitle%
\begin{abstract}
We study the problem of extracting a prescribed number of random bits by reading the smallest possible number of symbols from non-ideal stochastic processes. The related interval algorithm proposed by Han and Hoshi has asymptotically optimal performance; however, it assumes that the distribution of the input stochastic process is known. The motivation for our work is the fact that, in practice, sources of randomness have inherent correlations and are affected by measurement's noise. Namely, it is hard to obtain an accurate estimation of the distribution. This challenge was addressed by the concepts of seeded and seedless extractors that can handle general random sources with unknown distributions. However, known seeded and seedless extractors provide extraction efficiencies that are substantially smaller than Shannon's entropy limit. Our main contribution is the design of extractors that have a variable input-length and a fixed output length, are efficient in the consumption of symbols from the source, are capable of generating random bits from general stochastic processes and approach the information theoretic upper bound on efficiency.
\end{abstract}


\IEEEpeerreviewmaketitle

\section{Introduction}

\IEEEPARstart{W}{e}  study the problem of extracting a prescribed number of random bits by reading the smallest possible number of symbols from imperfect stochastic processes. For perfect stochastic processes, including processes with known accurate distributions or perfect biased coins, this problem has been well studied. It dates back to von Neumann [9] who considered the problem of generating random bits from a biased coin with unknown probability. Recently, in \cite{Zhou12_Streaming},  we improved von Neumann's scheme and introduced an algorithm that generates `random bit streams' from biased coins, uses bounded space and runs in expected linear time. This algorithm can generate a prescribed number of random bits with an asymptotically optimal efficiency. On the other hand, efficient algorithms have also been developed for extracting randomness from any known stochastic process (whose distribution is given). In \cite{Knuth1976}, Knuth and Yao presented a simple procedure for generating sequences with
arbitrary probability distributions from an unbiased coin (the probability of H and T is $\frac{1}{2}$). In \cite{Abrahams1996}, Abrahams considered
a source of biased coin whose distribution is an integer power of a noninteger.
Han and Hoshi \cite{Han1997}  studied the general problem and proposed an interval algorithm that generates a prescribed number of random bits from any known stochastic process and achieves the information-theoretic upper bound on efficiency.  However, in practice, sources of stochastic processes have inherent correlations and are affected by measurement's noise, hence, they are not perfect. Existing algorithms for extracting randomness from perfect stochastic processes cannot work for imperfect stochastic processes, where uncertainty exists.

To extract randomness from an imperfect stochastic process, one approach is to apply a seeded or seedless extractor to a sequence generated by the process that contains a sufficient amount of  randomness, and we call this approach as a fixed-length extractor for stochastic processes since all the possible input sequences have the same fixed length. Efficient constructions of seeded or seedless extractors have been extensively studied in last two decades, and it shows that the number of random bits extracted by them can approach the source's min-entropy asymptotically \cite{Dvir08, Nis96, Sha02,Rao2007,Kamp11}. Although fixed-length extractors can generate random bits with good quality from imperfect stochastic processes,
their efficiencies are not close to the optimality. Here, we define the \emph{efficiency} of an extractor for stochastic processes as the asymptotic ratio
between the number of extracted random bits and the entropy of its input sequence (the entropy of its input sequence is proportional to the expected input length if the stochastic process is stationary ergodic), which is upper bounded by $1$ since the process of extracting randomness does not increase entropy. Based on this definition, we can conclude that the efficiency of a fixed-length extractor is upper bounded by the ratio between the min-entropy and the entropy of the input sequence, which is usually several times smaller than $1$. So fixed-length extractors are not very efficient in extracting randomness from stochastic processes. The intuition is that, in order to minimize the expected number of symbols read from an imperfect stochastic process, the length of the input sequence should be adaptive, not being fixed.

The concept of min-entropy and entropy are defined as follows.
\begin{Definition}
Given a random source $X$ on $\{0,1\}^n$, the \emph{min-entropy} of $X$ is defined as
$$H_{\min}(X)=\min_{x\in \{0,1\}^n } \log \frac{1}{P[X=x]}.$$
The \emph{entropy} of $X$ is defined as
$$H(X)=\sum_{x\in \{0,1\}^n } P[X=x] \log \frac{1}{P[X=x]}.$$
\end{Definition}

The following example is constructed for comparing entropy with min-entropy for a simple random variable.
\begin{Example}
Let $X$ be a random variable such that $P[X=0]=0.9$ and $P[X=1]=0.1$, then
$H_{\min}(X)=0.152$ and $H(X)=0.469$. In this case, the entropy of $X$ is about three times its min-entropy.\hfill
	$\Box$
\end{Example}

In this paper, we focus on the notion and constructions of variable-length extractors (short for variable-to-fixed length extractors), namely, extractors with variable input length and fixed output length. (Note that the interval algorithm proposed by Han and Hoshi \cite{Han1997} and the streaming algorithm proposed by us \cite{Zhou12_Streaming} are special cases of variable-length extractors). Our goal is to extract a prescribed number of random bits in the sense of statistical distance while minimizing the expected input cost, measured by the entropy of the input sequence (whose length is variable). To make this precise, we let
$d(\mathcal{R},\mathcal{M})$ be the difference between two known stochastic processes $\mathcal{R}$ and $\mathcal{M}$, defined by
$$d(\mathcal{R},\mathcal{M})=\limsup_{n\rightarrow\infty} \max_{x\in \{0,1\}^n} \frac{\log_2\frac{P_{\mathcal{R}}(x)}{P_{\mathcal{M}}(x)}}{\log_2\frac{1}{P_{\mathcal{M}}(x)}},$$
where $P_{\mathcal{R}}(x)$ is the probability of generating $x$ from $\mathcal{R}$ when the sequence length is $|x|$, and $P_{\mathcal{M}}(x)$ is the probability of generating $x$ from $\mathcal{M}$ when the sequence length is $|x|$.

A few models of imperfect stochastic processes are introduced and investigated, including,
\begin{itemize}
  \item Let $\mathcal{M}$ be a known stochastic process, we consider an arbitrary stochastic process $\mathcal{R}$ such that $d(\mathcal{R},\mathcal{M})\leq \beta$ for a constant $\beta$.
  \item We consider $\mathcal{R}$ as an arbitrary stochastic process such that
  $\min_{\mathcal{M}\in \mathcal{G}_{s.e.}}d(\mathcal{R},\mathcal{M})\leq \beta$ for
  a constant $\beta$, where $\mathcal{G}_{s.e.}$ denotes the set consisting of all stationary ergodic processes.
\end{itemize}

Generally, given a real slight-unpredictable source $\mathcal{R}$, it is not easy to estimate the exact value of $d(\mathcal{R},\mathcal{M})$ for
a stochastic process $M$. But its upper bound, i.e., $\beta$, can be easily obtained.
The parameter
$\beta$ describes how unpredictable the real source $\mathcal{R}$ is, so we call it the \emph{uncertainty} of $\mathcal{R}$.
We prove that it is impossible to construct an extractor that achieves efficiency strictly larger than $1-\beta$
for all the possible sources $\mathcal{R}$ with uncertainty $\beta$. Then we introduce several constructions of variable-length extractors, and show that their efficiencies can reach $\eta \geq 1-\beta$; that is, the constructions are asymptotically optimal. The proposed variable-length extractors have two benefits: (i) they are generalizations of algorithms for perfect sources to  address general imperfect sources; and (ii) they bridge the gap between min-entropy and entropy on efficiency.

The following example is constructed to compare the performances of a variable-length extractor and
a fixed-length extractor when extracting randomness from a slightly-unpredictable independent process.
\begin{Example} Consider an independent process $x_1x_2x_3...$ such that $P[x_i=1]\in [0.9, 0.91]$, then it can be obtained that $\beta\leq 0.0315$.
For this source, a variable-length extractor can generate random bits with efficiency at least $1-\beta=0.9685$ that is very close to the upper bound $1$. In comparison, fixed-length extractors can only reach the efficiency at most $0.3117$.
\end{Example}

The remainder of this paper is organized as follows. Section \ref{sec_pre} presents background and related results.
In Section \ref{var_sec_efficiency}, we demonstrate that one cannot construct a variable-length extractor with efficiency strictly larger than $1-\beta$ when the source has uncertainty $\beta$. Then we focus on the seeded constructions of variable-length extractors, namely, we use a small number of additional truly random bits as the seed (catalyst).
Three different constructions are provided and analyzed in Section \ref{var_section_tech1}, Section \ref{var_section_tech2} and Section \ref{var_section_tech4}
separately. All these constructions have efficiencies lower bounded by $1-\beta$, implying their optimality.
Finally, we discuss seedless constructions of variable-length extractors for some types of random sources in Section \ref{sec_randomnessextraction}, followed by the concluding remarks.

\section{Preliminaries}
\label{sec_pre}

\subsection{Statistical Distance}

\emph{Statistical Distance} is used in computer science to measure the difference between two distributions. Let $X$ and $Y$ be two random sequences with range $\{0,1\}^m$, then the statistical distance between $X$ and $Y$ is defined as
$$\|X-Y\|=\max_{T:\{0,1\}^m\rightarrow \{0,1\}} |P[T(X)=1]-P[T(Y)=1]|$$
over a boolean function $T$.
We say that $X$ and $Y$ are $\epsilon$-close if $\|X-Y\|\leq \epsilon$. According to this definition, we can also write
$$\|X-Y\|=\frac{1}{2}\sum_{x\in \{0,1\}^m}|P[X=x]-P[Y=x]|\leq \epsilon.$$
It is equivalent to the former expression.

Let $U_m$ denote the uniform distribution on $\{0,1\}^m$. In order to let a sequence $Y$ to be able to take place of the truly random bits in a randomized application, we let $Y$ be $\epsilon$-close to $U_m$, where $\epsilon$ is small enough. In this case,
the extra probability error introduced by this replacement is at most $\epsilon$.
In this paper, we want to extract $m$ almost-random bits such that they form a sequence $\epsilon$-close to the uniform distribution $U_m$ on $\{0,1\}^m$ with specified small $\epsilon>0$, i.e.,
$$\|Y-U_m\|\leq \epsilon.$$

\subsection{Seeded Extractors}

In 1990, Zuckerman introduced a general model of weak random sources, called $k$-sources, namely whose min-entropy is at least $k$ \cite{Zuc90}.
It was shown that given a source on $\{0,1\}^n$ with min-entropy $k<n$, it is impossible to devise a single function that extracts even one bit of randomness. This observation led to the introduction of seeded extractors, which use a small number of additional truly random bits as the seed (catalyst). When simulating a probabilistic algorithm, one can simply eliminate the requirement of truly random bits by enumerating all possible strings for the seed and taking a majority vote on the final results. There are a variety of very efficient constructions of seeded extractors, summarized in \cite{Dvir08, Nis96, Sha02}. Mathematically, a seeded extractor
is a function,
$$E: \{0,1\}^n \times \{0,1\}^d \rightarrow \{0,1\}^m,$$
such that for every distribution $X$ on $\{0,1\}^n$ with $H_{\min}(X)\geq k$, the distribution $E(X,U_d)$ is $\epsilon$-close to
the uniform distribution $U_m$. Here, $d$ is the seed length, and we call such an extractor as a $(k,\epsilon)$ extractor.
There are a lot of works focusing on efficient constructions of seeded extractors. A standard application of the probabilistic method \cite{Radhakrishnan2000} shows that there exists a seeded extractor which can extract asymptotically $H_{\min}(X)$ random bits with $\log(n-H_{\min}(X))$ additional truly random bits. Recently, Guruswami, Umans and Vadhan \cite{Guruswami07} provided an explicit construction of seeded extractors, whose efficiency is  very close to the bound obtained based on the probabilistic method. Their main result is described as follows:

\begin{Lemma} \emph{\cite{Guruswami07}} \label{lemma_seededextractor} For every constant $\alpha>0$, and all positive integers $n, k$ and all $\epsilon>0$, there is an explicit construction of a $(k,\epsilon)$ extractor $E: \{0,1\}^n\times \{0,1\}^d\rightarrow\{0,1\}^m$ with $d\leq \log n+ O(\log(k/\epsilon))$ and $m\geq (1-\alpha)k$.
\end{Lemma}

The above result implies that given any source $X \in \{0,1\}^n$ with min-entropy $k$, if $\geq (1+\alpha)m$ with $\alpha>0$, we can always construct
a seeded extractor to generates a random sequence $Y\in \{0,1\}^m$ that is $\epsilon$-close to the uniform distribution. In this case,
the seed length $d\leq \log n + O(\log(k/\epsilon))$ depends on the input length $n$ and the parameter $\epsilon$.

\subsection{Seedless Extractors}

In the last decade, the concept of seedless (deterministic) extractors has attracted renewed interests, motivated by the reduction of the computational complexity for simulating probabilistic algorithms as well as some requirements in cryptography \cite{Dodis00}. Several specific classes of sources have been studied, including independent sources, which can be divided into several independent parts containing certain amount of randomness \cite{Barak06, Rao2007, Raz2005}; bit-fixing sources, where some bits in a binary sequence are truly random and the remaining bits are fixed \cite{Cohen89, Gabizon06, Kamp06}; samplable sources, where the source is generated by a process that has a bounded amount of computational resources like space \cite{Kamp11, Trevisan00}. For example, suppose that we have multiple independent sources with the same length $n$. It is known how to extract from two sources when
the min-entropy in each is $\geq 0.5n$ \cite{Raz2005} or slightly less than $0.5n$ \cite{Bou05}, how to extract from $O(1/\gamma)$ sources if the min-entropy in each is at least $n^\gamma$ \cite{Rao06}. All these constructions have exponentially small error, and they are able to extract $\Theta(k)$ random bits.

Both seeded extractors and seedless extractors described above have fixed input length, fixed seed length ($d=0$ for seedless extractors) and fixed output length. So we call them fixed-length extractors. To apply fixed-length extractors in extracting randomness from a stochastic process, it needs to first read a sequence of fixed length, whose min-entropy is strictly larger than the number of random bits that we need to generate. Fixed-length extractors can generate random bits of good quality from imperfect stochastic processes,
but they usually consume more incoming symbols than what are necessarily required. To increase information efficiency,
we let the length of input sequences be adaptive, hence, we have the concept of `variable-length extractors'.

\subsection{Variable-Length Extractors}

A variable-length extractor is an extractor with variable input length and fixed output length. When applying a variable-length extractor to
a stochastic process, it reads incoming symbols one by one until the whole incoming sequence meets certain criterion, then it maps the incoming sequence into a binary sequence of fixed length as the output. Depending on the sources, the construction may require a small number of additional truly random bits as the seed. Hence, we have seeded variable-length extractors and seedless variable-length extractors.

A seeded variable-length extractor is a function,
$$V_E: S_p\times\{0,1\}^d\rightarrow\{0,1\}^m,$$
such that given a real source $\mathcal{R}$, the output sequence is $\epsilon$-close to
the uniform distribution $U_m$. Here, $S_p$ is the set consisting of all possible input sequences, called the input set. It is complete and prefix-free. The input sequence is compete, that means, any infinite sequence has a prefix in the set; so when reading symbols from any source, we can always meet a sequence in the set. Then we stop reading and map this sequence into a binary sequence of length $m$. Being prefix-free is not very necessary; it ensures that all the sequences in $S_p$ are possible to read.

A general procedure of extracting randomness by using variable-length extractors can be divided into three steps:

\begin{enumerate}
  \item Determining an input set $S_p$ such that its min-entropy based on the real source $\mathcal{R}$ is at least $k$, namely,
$$\min_{x\in S_p}\log_2\frac{1}{P_\mathcal{R}(x)}\leq k,$$
where $k\geq (1+\alpha)m$ for any $\alpha>0$.
  \item We construct an injective function
$$V:S_p\rightarrow \{0,1\}^n,$$
to map the sequences in $S_p$ into binary sequences of length $m$. We read symbols from the source $\mathcal{R}$ one by one until
the current incoming sequence matches one in $S_p$. This incoming sequence is then mapped to a binary sequence of length $n$ based on function $V$. As a result, we get a random sequence $Z$ with length $n$ and min-entropy $k$ (since $V$ is injective).
  \item Since $k=(1+\alpha)$ with an $\alpha>0$, according to Lemma \ref{lemma_seededextractor}, we can always find a seeded extractor,
  $$E:\{0,1\}^n \times\{0,1\}^d\rightarrow\{0,1\}^m$$
  that can extract $m$ almost-random bits from a source with min-entropy $k$. By applying  this seeded extractor $E$ to the sequence $Z$,
  we get a random sequence of length $m$ that is $\epsilon$-close to the uniform distribution $U_m$. Here, the seed length $d\leq \log n + O(\log(k/\epsilon))$.
\end{enumerate}

We can see that the construction of a variable-length extractor is a cascade of
a function $V$ and a seeded extractor $E$, i.e.,
$$V_E=E\bigotimes V.$$

Note that our requirement is to extract a sequence of $m$ almost-random bits that is $\epsilon$-close to the uniform distribution $U_m$. The key of constructing variable-length extractors is to find the input set $S_p$ with min-entropy $k$, even the distribution of the real source $\mathcal{R}$ is slightly unpredictable, such that the expected length of the sequences in $S_p$ is minimized. For stationary ergodic processes, minimizing the expected length is equivalent to minimizing the entropy of the sequences in $S_p$ asymptotically (this will be discussed in this section).

For some specific types of sources, including independent sources and samplable sources, by applying the ideas in \cite{Rao2007} and \cite{Kamp11} we can remove
the requirement of truly random bits without degrading the asymptotic performance. As a result, we have seedless variable-length extractors.
For example, if the source $\mathcal{R}$ is an independent process, we can first apply the method in \cite{Rao2007} to extract $d$ almost-random bits from the first $\Theta(\log \frac{m}{\epsilon})$ bits, and then use them as the seed of a seeded variable-length extractor to extract randomness from the rest of the process. The detailed discussions will be given in Section \ref{sec_randomnessextraction}.

\section{Efficiency and Uncertainty}
\label{var_sec_efficiency}

\subsection{Efficiency}

To consider the performance of an extractor, we define its \emph{efficiency} as
the asymptotical ratio between the output length and the total entropy of all its inputs. So the efficiency of an extractor can be written as
$$\eta=\lim_{m\rightarrow\infty} \frac{m}{H_{\mathcal{R}}(X_m)+d},$$
such that the output sequence is $\epsilon$-close to the uniform distribution $U_m$
on $\{0,1\}^m$, where $\epsilon$ is small, $d$ is the seed length, $m$ is the output length, and $H_{\mathcal{R}}(X_m)$ is the entropy of the input sequence $X_m$ with range on $S_p$.
In our constructions,  $d\leq \log n + O(\log(m/\epsilon))$, which is ignorable compared to $H_{\mathcal{R}}(X_m)$ when $m\rightarrow\infty$. Hence,
we can write
$$\eta=\lim_{m\rightarrow\infty} \frac{m}{H_{\mathcal{R}}(X_m)}.$$
In the definition, we use the entropy of the input sequence rather than the expected input length, because the source that we considered may not be stationary ergodic.
It needs to mention that, in seeded constructions, the value of $d$ is also an important parameter although it is much smaller than $m$. The problem of minimizing the seed length $d$ can be studied separately from minimizing the entropy of the input sequence, and it will be addressed in this paper.

First, we demonstrate that if a distribution is $\epsilon$-close to the uniform distribution $U_m$, then
the entropy of this distribution is asymptotically $m$ for any $\epsilon<1$.

\begin{Lemma}\label{var_lemma2}
Let $X$ be a random sequence on $\{0,1\}^m$ that is $\epsilon$-close to the uniform distribution $U_m$, then
$$ m- \log_2\frac{1}{1-\epsilon}\leq H(X)\leq m.$$
\end{Lemma}

\proof Since there are totally $2^m$ possible assignments for $X$, it is easy to get $H(X)\leq m$.
So we only need to prove that
$$H(X)\geq m- \log_2\frac{1}{1-\epsilon}.$$

Let $p(x)$ denote $P[X=x]$ for $x\in \{0,1\}^m$.
Since $X$ is  $\epsilon$-close to the uniform distribution $U_m$, we have
$$\frac{1}{2} \sum_{x\in \{0,1\}^m} \|p(x)-2^{-m}\|\leq \epsilon.$$

Then the lower bound of $H(X)$ can be written as
$$\min_{p} \sum_{x\in \{0,1\}^m} p(x) \log_2\frac{1}{p(x)}$$
subject to
$$p(x)\geq 0, \forall x\in \{0,1\}^m;$$
$$\sum_{x\in\{0,1\}^m}p(x)=1;$$
$$\sum_{x\in \{0,1\}^m} \|p(x)-2^{-m}\|\leq 2\epsilon.$$

Obviously, the optimal solution of the above  problem happens at
$$\sum_{x\in \{0,1\}^m} \|p(x)-2^{-m}\|=2\epsilon.$$

To solve the problem based on Lagrange Multipliers, we let
$$\lambda(p)= \sum_{x\in \{0,1\}^m}\ p(x) \log_2\frac{1}{p(x)}+\lambda_1(\sum_{x\in\{0,1\}^m}p(x)-1)$$
$$+\lambda_2 (\sum_{x\in \{0,1\}^m} \|p(x)-2^{-m}\|-2\epsilon).$$

If $p(x)\geq 0$ with $x\in\{0,1\}^m$ is a solution of the above question, then
$$\frac{\partial \lambda }{\partial(p(x))}=0,$$
i.e.,
$$ \left\{ \begin{array}{cc}
          \frac{\ln p(x)+1}{\ln 2}+ \lambda_1+\lambda_2=0    &  \textrm{ if } 2^{-m}\leq p(x)
          \leq 1, \\
           \frac{\ln p(x)+1}{\ln 2}+ \lambda_1-\lambda_2=0   &  \textrm{ if } 0\leq p(x)\leq 2^{-m}.
           \end{array}
\right.$$

So there exists two constants $a$ and $b$ with $0\leq a\leq 2^{-m}\leq b\leq 1$, such that,
$$ \left\{ \begin{array}{cc}
          p(x)=a   &  \textrm{ if } 2^{-m}\leq p(x)\leq 1, \\
          p(x)=b   &  \textrm{ if } 0\leq p(x)\leq 2^{-m}.
           \end{array}
\right.$$

Assume that there are $t$ assignments of $x$ with $p(x)=a$, then there are $2^m-t$ assignments of $x$ with $p(x)=b$.
Hence, the problem is converted to the one over $a,b,t$, i.e.,
$$\min_{a,b,t} t a\log \frac{1}{a} +(2^m-t) b\log \frac{1}{b},$$
subject to
$$0\leq t\leq 2^m;$$
\begin{equation}ta +(2^m-t)b=1;\label{var_equ_entropy1}\end{equation}
\begin{equation}t(2^{-m}-a)+(2^m-t)(b-2^{-m})=2\epsilon.\label{var_equ_entropy2}\end{equation}
From Equ.~(\ref{var_equ_entropy1}) and (\ref{var_equ_entropy2}), we get
$$a=2^{-m}-\frac{\epsilon}{t}, \quad b=2^{-m}+\frac{\epsilon}{2^m-t}.$$

So the question is finding the optimal $t$ that minimizes
$$-t(2^{-m}-\frac{\epsilon}{t})\log_2 (2^{-m}-\frac{\epsilon}{t})$$
$$-(2^m-t)(2^{-m}+\frac{\epsilon}{2^m-t})\log_2(2^{-m}+\frac{\epsilon}{2^m-t}),$$
subject to
$$0\leq t\leq \frac{\epsilon}{2^{-m}}.$$

The optimal solution is $t^*=\frac{\epsilon}{2^{-m}}$. In this case, the entropy of $X$ is
$$H(X)= \log (2^m-t)=m-\log_2\frac{1}{1-\epsilon},$$
which is the lower bound.

This completes the proof.
\hfill\QED

In the following lemma, we show that for any extractor, its efficiency is upper bounded by $1$. The reason is that the amount of information, i.e., entropy, does not increase
during the process of randomness extraction.

\begin{Lemma}\label{lemma_efficiency1}
For any extractor with seed length $d$ and output length $m$, if $d=o(m)$, its efficiency $\eta\leq 1$.
\end{Lemma}

\proof We consider fixed-length extractors as a special case of variable-length extractors, and consider seedless extractors as a special case of seeded extractors when $d=0$. So  our proof only focus on seeded variable-length extractors.

A main observation is that for any extractor, the entropy of its output sequence is bounded
by the entropy of the input sequence plus the entropy of the seed, since the process of extracting randomness cannot create new randomness.

For the output sequence, denoted by $Y$, it is $\epsilon$-close to the uniform distribution $U_m$. According to
Lemma \ref{var_lemma2}, its entropy is $$H_{\mathcal{R}}(Y)\geq m-\log_2\frac{1}{1-\epsilon}.$$

The total entropy of the inputs is $H_{\mathcal{R}}(X_m)+d$. Hence,
$$H_{\mathcal{R}}(Y)\leq H_{\mathcal{R}}(X_m)+d.$$

As a result, the efficiency of the extractor is
$$\eta=\lim_{m\rightarrow\infty} \frac{m}{H_{\mathcal{R}}(X_m)}=\lim_{m\rightarrow\infty}\frac{H_{\mathcal{R}}(Y)}{H_{\mathcal{R}}(X_m)+d}\leq 1.$$

This completes the proof.
\hfill\QED

If $\mathcal{R}$ is a stationary ergodic process, we define its entropy rate as
$$h(\mathcal{R})=\lim_{l\rightarrow\infty}\frac{H(X^l)}{l},$$
where $X^l$ is a random sequence of length $l$ generated from the source $\mathcal{R}$.
In this case, the entropy of the input sequence on $S_p$ is proportional to the expected input length.

\begin{Lemma} \label{var_theorem_stionaryergodiclength} Given a stationary ergodic source $\mathcal{R}$, let $X_m$ be the input sequence of a variable-length extractor that has an output length $m$. Then
$$\lim_{m\rightarrow\infty} \frac{H_{\mathcal{R}}(X_m)}{E_{\mathcal{R}}[|X_m|]}=h(\mathcal{R}),$$
where $E_{\mathcal{R}}[|X_m|]$ is the expected input length.
\end{Lemma}

\proof $X_m$ is a random sequence from $S_p$ based on the distribution of $\mathcal{R}$.
Let $l_1$ be the minimum length of the sequences in $S_p$, as $m\rightarrow\infty$, $l_1\rightarrow\infty$.
Now, we define $$l_i=l_1+(i-1)\log l_1 \textrm{ for all }i\geq 1.$$ Based on them, we divide all the sequences in $S_p$ into subsets
$$S_i=\{x|x\in S_p, l_i\leq |x|\leq l_{i+1}-1\}$$
for $i\geq 1$.

Let $p_i=P_{\mathcal{R}}(X_m\in S_i)$, then
$$H_{\mathcal{R}}(X_m)
   \geq \sum_{i}[(\sum_{j>i} p_j)H_{\mathcal{R}}(X_{l_{i-1}+1}^{l_i}|X_{1}^{l_{i-1}}, |X_m|\geq l_i)],$$
where $l_0=0$, $\sum_{j>i} p_j$ is the probability that $|X_m|\geq l_i$, and
$X_a^b$ is a sequence of $X_m$ from the $a$th element to the $b$th element.

Since $X_m$ is generated from a stationary ergodic process, and $l_i-{l_{i-1}}\rightarrow\infty$ as $m\rightarrow\infty$, we can get
$$H_{\mathcal{R}}(X_{l_{i-1}+1}^{l_i}|X_{1}^{l_{i-1}},  |X_m|\geq l_i)\rightarrow (l_i-l_{i-1})h(\mathcal{R}).$$

As a result, as $l_1\rightarrow\infty$, we have
\begin{eqnarray*}
H_{\mathcal{R}}(X_m)&\geq& (1-\epsilon) \sum_{i}(\sum_{j>i} p_j) (l_i-l_{i-1})h(\mathcal{R})\\
&=&(1-\epsilon)\sum_i p_i l_i h(\mathcal{R}),
\end{eqnarray*}
for an arbitrary $\epsilon>0$.

Also considering the other direction, we can get
that as $l_1\rightarrow\infty$,
\begin{eqnarray*}
 H_{\mathcal{R}}(X_m)
&\leq &(1+\epsilon)\sum_i p_i l_{i+1} h(\mathcal{R})\\
&=& (1+\epsilon)\sum_i p_i (l_i+\log l_1) h(\mathcal{R}),
\end{eqnarray*}
for an arbitrary $\epsilon>0$.

For the expected input length, i.e., $E_{\mathcal{R}}[|X_m|]$, it is easy to show that
$$\sum_i p_i l_i\leq E_{\mathcal{R}}[|X_m| ] \leq \sum_i p_i l_{i+1}=\sum_i p_i( l_i+\log l_1)  .$$

So as $m\rightarrow \infty$, i.e., $l_1\rightarrow\infty$, it yields
$$\lim_{m\rightarrow\infty} \frac{H_{\mathcal{R}}(X_m)}{E_\mathcal{R}[|X_m|]}=\lim_{m\rightarrow\infty}\frac{\sum_i p_i l_i h(\mathcal{R})}{\sum_i p_i l_i}$$
$$=h(\mathcal{R}).$$

This completes the proof.
\hfill\QED

\subsection{Sources and Uncertainty}

Given a source $\mathcal{R}$, if its distribution is known, we say that this source is a known stochastic process, and its uncertainty is $0$. In this paper,
we mainly focus on those imperfect processes whose distributions are slightly unpredictable due to many factors like
the existence of external adversaries.

First, given two known stochastic processes $\mathcal{R}$ and $\mathcal{M}$, we let
$d(\mathcal{R},\mathcal{M})$ be the difference between $\mathcal{R}$ and $\mathcal{M}$.  Here, we define $d(\mathcal{R},\mathcal{M})$ as
$$d(\mathcal{R},\mathcal{M})=\limsup_{n\rightarrow\infty} \max_{x\in \{0,1\}^n} \frac{\log_2\frac{P_{\mathcal{R}}(x)}{P_{\mathcal{M}}(x)}}{\log_2\frac{1}{P_{\mathcal{M}}(x)}},$$
where $P_{\mathcal{R}}(x)$ is the probability of generating $x$ from $\mathcal{R}$ when the sequence length is $|x|$, and $P_{\mathcal{M}}(x)$ is the probability of generating $x$ from $\mathcal{M}$ when the sequence length is $|x|$. Although there are some existing ways such as normalized Kullback-Leibler divergence to measure the difference between two sources, with them it is not easy to estimate the uncertainty of a source and it is
not easy to analyze the performances of constructed variable-length extractors.

In the rest of this paper, we investigate a few models of unpredictable sources. Most natural source can be well described in those ways.

\begin{enumerate}
  \item The source $\mathcal{R}$ is an arbitrary stochastic process such that $d(\mathcal{R},\mathcal{M})\leq \beta$ for a constant $\beta\in [0,1]$ and
  a known stochastic process $\mathcal{M}$.
  \item $\mathcal{R}$ is an arbitrary stochastic process such that there exists a stationary ergodic process $\mathcal{M}$ (whose distribution is unknown)
and $d(\mathcal{R},\mathcal{M})\leq \beta$; that is, $\min_{\mathcal{M}\in \mathcal{G}_{s.e.}}d(\mathcal{R},\mathcal{M})\leq \beta$,  where $\mathcal{G}_{s.e.}$ denotes the set consisting of all stationary ergodic processes.
\end{enumerate}

In both the models, we call $\beta$ as the \emph{uncertainty} of the source $\mathcal{R}$. In the real world,
$\beta$ can be easily estimated without knowing the distribution of the processes.
It just reflects how unpredictable the real source $\mathcal{R}$ is.

To construct variable-length extractors, we only care about the possible input sequences, namely, those in $S_p$. Hence, for the case
of finite length, $d_p(\mathcal{R},\mathcal{M})$ is a more important parameter for us, defined by
$$d_p(\mathcal{R},\mathcal{M})= \max_{x\in S_p} \frac{\log_2\frac{P_{\mathcal{R}}(x)}{P_{\mathcal{M}}(x)}}{\log_2\frac{1}{P_{\mathcal{M}}(x)}},$$

As the number of required random bits $m$ increases, $d_p(\mathcal{R},\mathcal{M})$ quickly converge to $d(\mathcal{R}, \mathcal{M})$.
And we can write
$$d_p(\mathcal{R},\mathcal{M})=d(\mathcal{R}, \mathcal{M})+\epsilon_p$$
for a very small constant $\epsilon_p$. As $m\rightarrow\infty$, $\epsilon_p\rightarrow 0$. In this case, the upper bound of $d_p(\mathcal{R},\mathcal{M})$ or
$\min_{\mathcal{M}\in \mathcal{G}_{s.e.}}d_p(\mathcal{R},\mathcal{M})$ is
$$\beta_p=\beta+\epsilon_p.$$

\begin{Example} Let $x_1x_2...\in \{0,1\}^*$ be a sequence generated from an independent source $\mathcal{R}$ such that $$\forall i\geq 1, P[x_i=1]\in [0.8,0.82].$$
If we let $\mathcal{M}$ be a biased coin with probability $0.8132$, then
$$\beta=\max_{\textrm{possible }\mathcal{R}} d(\mathcal{R},\mathcal{M})$$
$$=\max(\frac{\log_2 \frac{0.2}{0.1868}}{\log_2 \frac{1}{0.1868}}, \frac{\log_2 \frac{0.82}{0.8132}}{\log_2 \frac{1}{0.8132}})=0.0405.$$ \hfill  $\Box$
\end{Example}

According to our definition, $d(\mathcal{M},\mathcal{R})\leq \beta$ if and only if
$$P_{\mathcal{R}}(x)\leq P_{\mathcal{M}}(x)^{1-\beta}$$
for all $x\in\{0,1\}^\infty$ with  $|x|\rightarrow\infty$. This is a condition that is very easy to be satisfied by many natural stochastic processes for a small $\beta$.

\begin{Lemma}  If $d(\mathcal{R},\mathcal{M})\rightarrow 0$, we have
$$P_{\mathcal{R}}(x)\rightarrow P_{\mathcal{M}}(x) $$
for all $x\in \{0,1\}^*$.
\end{Lemma}

\subsection{Efficiency and Uncertainty}

In this subsection, we investigate the relation between the efficiency and uncertainty. We show that given
a stochastic process $\mathcal{R}$ with uncertainty $\beta$, as described in the previous subsection, one cannot construct a variable-length extractor with efficiency strictly larger than $1-\beta$ for all the possibilities of $\mathcal{R}$.

Let us first consider a simple example: let $X$ be a random sequence with the uniform distribution on $\{0,1\}^n$ and let $Y$
be an arbitrary random sequence on $\{0,1\}^n$ such that $$\frac{\log_2 \frac{ P[Y=x]}{P[X=x]}}{\log_2\frac{1}{P[X=x]}}\leq \beta, \forall x\in \{0,1\}^n.$$
Now, we show that from the source $Y$, one cannot construct an extractor with efficiency strictly larger than $1-\beta$.
To see this, we consider an extractor $f$ with output length $m$, and a source $Y$ with
$$P[Y=y]\in \{0, 2^{-n(1-\beta)}\}, \forall y\in \{0,1\}^n.$$
For this a source $Y$, its entropy is $H(Y)=n(1-\beta)$. In order to make sure the output sequence of $f$, denoted by $Z$,
is $\epsilon$-close to $U_m$, it has
$$\lim_{m\rightarrow\infty} \frac{m}{n(1-\beta)}\leq \lim_{m\rightarrow\infty} \frac{H(Z)+\log_2\frac{1}{1-\epsilon}}{H(Y)}\leq 1.$$
So we cannot generate more than $n(1-\beta)$ random bits asymptotically. In this case, if we apply the seeded extractor $f$
to the random sequence $X$, which is a possibility of $Y$, then the efficiency is
$$\eta=\lim_{m\rightarrow\infty}\frac{m}{H(X)}=\lim_{m\rightarrow\infty}\frac{m}{n}\leq 1-\beta.$$
So there does not exist a seeded extractor that can extract randomness from an arbitrary $Y$ and its efficiency is strictly larger than $1-\beta$.
Here, $\beta$ is the uncertainty of the source.

\begin{Theorem} \label{var_theorem_lowerbound1} Let $\mathcal{M}$ be a known stochastic process, and $\mathcal{R}$ be an arbitrary stochastic process such that
$d(\mathcal{R},\mathcal{M})\leq \beta$, then one cannot construct a variable-length extractor whose efficiency
is strictly larger than $1-\beta$ for all possible $\mathcal{R}$.
\end{Theorem}

\proof Let $f$ be a variable-length extractor whose input sequence is a random sequence $X_m$ on $S_p$ and
its output sequence is a random sequence $Y$ on $\{0,1\}^m$. Assume that as $m\rightarrow\infty$, $f$ can extract from an arbitrary $\mathcal{R}$ such that the output sequence  $Y$ is $\epsilon$-close to $U_m$.

Let $h=H_{\mathcal{M}}(X_m)$ be the entropy of the input sequence based on the distribution of $\mathcal{M}$,
then we want to show that there exists a process $\mathcal{R}$ such that $d(\mathcal{R},\mathcal{M})\leq \beta$ and
$H_{\mathcal{R}}(X_m)\leq h(1-\beta)$ as $m\rightarrow\infty$.

To find such a process $\mathcal{R}$, we order all the elements in $S_p$ as
$x_1, x_2, x_3, ...$
such that
$$P_{\mathcal{M}}(x_1)\geq P_{\mathcal{M}}(x_2)\geq P_{\mathcal{M}}(x_3)\geq ...$$

Then we divide all these elements into groups,
$$\{x_1, x_2, ..., x_{i_1}\}, \{x_{i_1+1},x_{i_1+2},...,x_{i_2}\},...$$
such that the total probability of the elements in each group is almost the probability of its first element to the power of $1-\beta$, i.e.,
$$0\leq P_{\mathcal{M}}(x_{i_j+1})^{1-\beta}-\sum_{k=i_j+1}^{i_{j+1}} P_{\mathcal{M}}(x_k)\leq P_{\mathcal{M}}(x_{i_j+1}),$$  for all  $j\geq 0$,
where $i_0=0$.

Let $A=\{x_{1}, x_{i_1+1}, x_{i_2+1}, ...\}$ be the set consisting of the first elements of all the groups.
Now, we consider a possibility of $\mathcal{R}$ in the following way: for all $x\in \{x_{1}, x_{i_1+1}, x_{i_2+1}, ...\}$,
its probability is
$$P_{\mathcal{R}}(x)=\sum_{k=i_j+1}^{i_{j+1}} P_{\mathcal{M}}(x_k), \textrm{ if } x=x_{i_j+1};$$
For all $x\in S_p/A=S_p/\{x_{1}, x_{i_1+1}, x_{i_2+1}, ...\}$, its probability is
$$P_{\mathcal{R}}(x)=0.$$

For this source $\mathcal{R}$, the entropy of the input sequence is
$$H_{\mathcal{R}}(X_m)=\sum_{x\in S_p} P_{\mathcal{R}}(x) \log_2 \frac{1}{P_{\mathcal{R}}(x)}.$$

As $m\rightarrow\infty$, we have
\begin{eqnarray*}
&&H_{\mathcal{R}}(X_m)\\
   &= & \sum_{x\in A} P_{\mathcal{R}}(x) \log_2 \frac{1}{P_{\mathcal{R}}(x)}  \\
 &\rightarrow& (1-\beta) \sum_{x\in A}  P_{\mathcal{R}}(x) \log_2 \frac{1}{P_{\mathcal{M}}(x)}\\
&=& (1-\beta)  \sum_{j\geq 0} \sum_{k=i_j+1}^{i_{j+1}} P_{\mathcal{M}}(x_k) \log_2 \frac{1}{P_{\mathcal{M}}(x_{i_j+1})}\\
&\leq & (1-\beta)  \sum_{j\geq 0}  \sum_{k=i_j+1}^{i_{j+1}} P_{\mathcal{M}}(x_k) \log_2 \frac{1}{P_{\mathcal{M}}(x_k)}\\
&=& (1-\beta) H_{\mathcal{M}}(X_m)\\
&=& (1-\beta)h.
\end{eqnarray*}

According to Lemma \ref{var_lemma2}, as $m\rightarrow\infty$, $\frac{m}{H_{\mathcal{R}}(Y)}\rightarrow 1$.
Furthermore, we can get $$\lim_{m\rightarrow\infty}\frac{H_{\mathcal{R}}(Y)}{H_{\mathcal{R}}(X_m)} \leq 1, $$
it implies that
$$\lim_{m\rightarrow\infty} \frac{m}{(1-\beta)h}\leq 1,$$
otherwise, the output sequence cannot be $\epsilon$-close to the uniform distribution $U_m$.

If we apply the extractor $f$ to the source $\mathcal{M}$, which is also a possibility for $\mathcal{R}$, then its efficiency is
$$\eta=\lim_{m\rightarrow\infty}\frac{m}{h}\leq 1-\beta.$$

So it is impossible to construct a variable-length extractor with efficiency strictly larger than $1-\beta$ for all the possibilities of the source $\mathcal{R}$. This completes the proof.
\hfill\QED

With the same proof, we can also get the following theorem.

\begin{Theorem} Let $\mathcal{R}$ be an arbitrary stochastic process such that
$d(\mathcal{R},\mathcal{M})\leq \beta$ for a stationary ergodic process $\mathcal{M}$ with unknown distribution,
, then one cannot construct a variable-length extractor whose efficiency
is strictly larger than $1-\beta$ for all possible $\mathcal{R}$.
\end{Theorem}

The above theorems show that one cannot construct an extractor whose efficiency is strictly larger than $1-\beta$ for all the possible source $\mathcal{R}$. Here, $\beta$ is an important parameter that measures the uncertainty of a real source $\mathcal{R}$, either to
a known process or to the nearest stationary ergodic process.
In the next a few sections, we will present a few constructions for efficiently extracting randomness from the sources described in this section.
We show that their efficiency $\eta$ satisfies
$$1-\beta\leq \eta\leq 1.$$
That means the bound $1-\beta$ is actually achievable and the constructions proposed in this paper are asymptotically optimal on efficiency.

\section{Construction I: Approximated by Known Processes}
\label{var_section_tech1}

In this section, we consider those sources which can be approximated by a known stochastic process $\mathcal{M}$, namely, an arbitrary process
$\mathcal{R}$ with $d(\mathcal{R},\mathcal{M})\leq \beta$ for a known process $\mathcal{M}$.
We say that a stochastic process $\mathcal{M}$ is known if its distribution is given, i.e., $P_{\mathcal{M}}(x)$ can be easily calculated for any $x\in \{0,1\}^*$.
Note that this process $\mathcal{M}$ is not necessary to be stationary or ergodic. For instance, $\mathcal{M}$ can be an independent process
$z_1z_2...\in \{0,1\}^*$ such that $$\forall i\geq1, P_{\mathcal{M}}(z_i=1)= \frac{1+sin(i/10)}{2}.$$

\subsection{Construction}

Our goal is to extract randomness from an imperfect random source $\mathcal{R}$. The problem is that
we do not know the exact distribution of $\mathcal{R}$, but we know that it can be approximated by a known process $\mathcal{M}$.
So we can use the distribution of $\mathcal{M}$ to estimate the distribution of $\mathcal{R}$. As a result, we have the following procedure to extract $m$ almost-random bits.

The idea of the procedure is first producing a random sequence of length $n$ and min-entropy $k=m(1+\alpha)$ with $\alpha>0$, from which
we can further obtain a sequence $\epsilon$-close to the uniform distribution $U_m$ by applying a $(k,\epsilon)$  seeded extractor.
According to the results of seeded extractors, this constant $\alpha>0$ can be arbitrarily small.

\begin{Construction}\label{const:1}
Assume the real source $\mathcal{R}$ is an arbitrary stochastic process such that $d(\mathcal{R}, \mathcal{M})\leq \beta$
for a known process $\mathcal{M}$. Then we extract $m$ almost-random bits from $\mathcal{R}$ based on the following procedure.

\begin{enumerate}
  \item Read input bits one by one from $\mathcal{R}$ until we get an input sequence $x\in \{0,1\}^*$ such that
  $$\log_2 \frac{1}{P_{\mathcal{M}}(x)}\geq \frac{k}{1-\beta_p},$$
  where  $\beta_p=\beta+\epsilon_p$ with $\epsilon_p>0$ and $k=m(1+\alpha)$ with $\alpha>0$. The small constant $\epsilon_p$ has value depending on the input set $S_p$; as
  $m\rightarrow\infty$, $\epsilon_p\rightarrow 0$. The constant $\alpha$ can be arbitrarily small.
  \item  Let $n$ be the maximum length of all the possible input sequences, then
  {$$n=\arg\min_{l}\{l\in \mathbb{N}|\forall y\in \{0,1\}^l, $$
  $$\log_2\frac{1}{P_\mathcal{M}(y)}\geq \frac{k}{1-\beta_p}\}.$$}
  If $|x|<n$, we  extend the length of $x$ to $n$ by adding $n-|x|$ trivial zeros at the end.
 Since $x$ is randomly generated, from the above procedure we get a random sequence $Z$ of length $n$. And it can be proved
 that this random sequence has min-entropy $k$.
  \item Applying a $(k,\epsilon)$ extractor to $Z$ yields a binary sequence of length $m$ that is $\epsilon$-close to the uniform distribution $U_m$.  \hfill$\Box$
\end{enumerate}
\end{Construction}

The following example is provided for comparing this construction with fixed-length constructions.

\begin{Example} Let $\mathcal{M}$ be a biased coin with probability $0.8$ (of being $1$). If $\frac{k}{1-\beta_p}=2$, then we can get the input set
$$S_p=\{0, 10, 110, 1110, 11110,111110, 1111110, 1111111\}.$$
In this case, the expected input length is strictly smaller than $7$.
For fixed-length constructions, to get a random sequence with min-entropy at least $2$, we have to read $7$ input bits independent of the context.
It is less efficient than the former method.\hfill
	$\Box$
\end{Example}

\begin{Theorem} Construction \ref{const:1} generates a random sequence of length $m$ that is $\epsilon$-close to $U_m$.
\end{Theorem}

\proof We only need to prove that given a source $\mathcal{R}$ and a model $\mathcal{M}$ with $d_p(\mathcal{R},\mathcal{M})\leq \beta_p$, it yields a random sequence $Z$ with min-entropy at least $k$.

According to the definition of $d_p(\mathcal{R},\mathcal{M})$, for all $x\in S_p$,
$$\frac{\log_2\frac{P_\mathcal{R}(x)}{P_\mathcal{M}(x)}}{\log_2 \frac{1}{P_\mathcal{M}(x)}}\leq \beta_p.$$

Based on the construction, for all $x\in S_p$
$$\log_2\frac{1}{P_\mathcal{M}(x)}\geq \frac{k}{1-\beta_p}.$$

The two inequalities above yield that
$$\log_2 \frac{1}{P_\mathcal{R}(x)}\geq k,$$
for all $x\in S_p$.

Since the second step, i.e., adding trivial zeros, does not reduce the min-entropy of $S_p$. As
a result, we get a random sequence $Z$ of length $n$ and with min-entropy at least $k$.

Since $k=m(1+\alpha)$ with $\alpha>0$, according to Lemma \ref{lemma_seededextractor},
we can construct a seeded extractor that applies to the sequence $Z$ and generates a binary sequence $\epsilon$-close to the uniform distribution $U_m$.

This completes the proof.
\hfill\QED

\subsection{Efficiency Analysis}

Now, we study the efficiency of Construction \ref{const:1}. According to our definition, given
a construction, its efficiency is
$$\eta=\lim_{m\rightarrow\infty}\frac{m}{H_{\mathcal{R}}(X_m)}.$$

\begin{Theorem}\label{var_theorem1_2} Given a real source $\mathcal{R}$ and a known process $\mathcal{M}$ such that $d(\mathcal{R},\mathcal{M})\leq \beta$, then the efficiency of Construction \ref{const:1} is
$$1-\beta\leq \eta\leq 1.$$
\end{Theorem}

\proof Since $\eta$ is always upper bounded by $1$, we only need to show that $\eta\geq 1-\beta$.

 According to Lemma \ref{lemma_seededextractor}, as $m\rightarrow \infty$,
we have
$$\lim_{m\rightarrow\infty}\frac{k}{m}=1.$$

Now, let us consider the number of elements in $S_p$, i.e., $|S_p|$. To calculate $|S_p|$, we let
$$S_p'=\{x[1:|x|-1] |x\in S_p\},$$
where $x[1:|x|-1]$ is the prefix of $x$ of length $|x|-1$,
then for all $y\in S_p'$, $$\log_2\frac{1}{P_\mathcal{M}(y)}\leq \frac{k}{1-\beta_p}.$$  Hence,
$$\log_2|S_p'|\leq \frac{k}{1-\beta_p}.$$

It is easy to see that $|S_p|\leq 2|S_p'|$, so
$$\log_2|S_p|\leq \frac{k}{1-\beta_p}+1.$$

Let $X_m$ be the input sequence, then
$$\lim_{k\rightarrow\infty}\frac{H_{\mathcal{R}}(X_m)}{k}\leq \lim_{k\rightarrow\infty}\frac{\log_2|S_p|}{k}$$
$$\leq \lim_{k\rightarrow\infty}\frac{1}{1-\beta_p}=\frac{1}{1-\beta}.$$

Finally, it yields
$$\eta=\lim_{m\rightarrow\infty}\frac{m}{H_{\mathcal{R}}(X_m)}\geq 1-\beta.$$

This completes the proof.\hfill\QED

We see that the efficiency of the above construction is between $1-\beta$ and $1$. As shown in Theorem \ref{var_theorem_lowerbound1}, the
gap $\beta$, introduced by the uncertainty of the real source $\mathcal{R}$, cannot be smaller.  Our construction is asymptotically optimal in the sense that we cannot find a variable-length extractor with efficiency definitely larger than $1-\beta$.

\begin{Corollary} Given a real source $\mathcal{R}$ and a known process $\mathcal{M}$ such that $d(\mathcal{R},\mathcal{M})\leq \beta$, then as $\beta\rightarrow 0$, the efficiency of Construction \ref{const:1} is
$$\eta\rightarrow 1.$$
\end{Corollary}

In this case, the efficiency of the construction can achieve Shannon's limit.

If $\mathcal{R}$ is a stationary ergodic process, we can also get the following result.

\begin{Corollary}  Given a stationary ergodic process$\mathcal{R}$ and a known process $\mathcal{M}$ such that $d(\mathcal{R},\mathcal{M})\leq \beta$, for the expected input length of Construction \ref{const:1}, we have
$$\frac{1}{h(\mathcal{R})}\leq \lim_{m\rightarrow\infty} \frac{E[|X_m|]}{m}\leq \frac{1}{(1-\beta)h(\mathcal{R})},$$
where $h(\mathcal{R})$ is the entropy rate of the source $\mathcal{R}$.
\end{Corollary}

\proof This conclusion is immediate following Lemma \ref{var_theorem_stionaryergodiclength}  and Theorem \ref{var_theorem1_2}.
\hfill
\QED

\section{Construction II: Approximately Biased Coins}
\label{var_section_tech2}

In this section, we use a general ideal model such as a biased coin or a Markov chain to approximate the real source $\mathcal{R}$.
Here, we do not care about the specific parameters of the ideal model. The reason is, in some cases, the source $\mathcal{R}$ is
very close to an ideal source but we cannot (or do not want to) estimate the parameters accurately.
As a result, we introduce a construction by exploring the characters of biased coins or Markov chains. For simplicity,
we only discuss the case that
the ideal model is a biased coin, and the same idea can be generalized when the ideal model is a Markov chain. Specifically,
let $\mathcal{G}_{b.c.}$ denote the set consisting of all the models of biased coins with different probabilities, and we
consider $\mathcal{R}$ as an arbitrary stochastic process such that
$$\min_{\mathcal{M}\in \mathcal{G}_{b.c.}} d(\mathcal{R},\mathcal{M}) \leq \beta.$$

\subsection{Construction}

The idea of the construction is similar as Construction \ref{const:1}, i.e., we  first produce a random sequence of length $n$ and with min-entropy $k=m(1+\alpha)$ for $\alpha>0$, from which
we can further obtain a sequence $\epsilon$-close to the uniform distribution $U_m$ by applying a $(k,\epsilon)$  seeded extractor.

\begin{Construction}\label{const:2}
Assume the real source $\mathcal{R}$ is an arbitrary stochastic process such that $$\min_{\mathcal{M}\in \mathcal{G}_{b.c.}} d(\mathcal{R},\mathcal{M}) \leq \beta$$ for a constant $\beta$.   Then we extract $m$ almost-random bits from $\mathcal{R}$ based on the following procedure.
\begin{enumerate}
\item Read input bits one by one from $\mathcal{R}$ until we get an input sequence $x\in \{0,1\}^*$ such that
     $$\log_2 {\nchoosek{k_0+k_1}{ \max(1,\min(k_0,k_1))}}\geq \frac{k}{1-\beta_p},$$
     where $k_0$ is the number of zeros in $x$, $k_1$ is the number of ones in $x$,
    $\beta_p=\beta+\epsilon_p$ with $\epsilon_p>0$ and $k=m(1+\alpha)$ with $\alpha>0$. The small constant $\epsilon_p$ has value depending on the input set $S_p$; as
  $m\rightarrow\infty$, $\epsilon_p\rightarrow 0$. The constant $\alpha$ can be arbitrarily small.
\item Since the input sequence $x$ can be very long, we map it into a sequence $z$ of fixed length $n$ such that
$$z=[I_{(k_0\geq k_1)}, \min(k_0,k_1), r(x)],$$
where $I_{(k_0\geq k_1)}=1$ if and only if $k_0\geq k_1$, and $r(x)$ is the rank of $x$  among all the permutations of $x$ with respect to the lexicographic order.
Since $x$ is randomly generated, the above procedure leads us to a random sequence $Z$ of length $n$.
\item Applying a $(k,\epsilon)$ extractor to $Z$ yields a random sequence of length $m$ that is $\epsilon$-close to $U_m$.\hfill$\Box$
\end{enumerate}
\end{Construction}

To see that the construction above works, we need to show that the random sequence $Z$ obtained after the second step has min-entropy at least $k$, and its length $n$ is well bounded.

\begin{Lemma}
 Given a source $\mathcal{R}$ with $\min_{\mathcal{M}\in \mathcal{G}_{b.c.}} d(\mathcal{R},\mathcal{M}) \leq \beta$, Construction \ref{const:2} yields a random sequence $Z$ with length
 $$n\leq 1+\lceil\log_2(\frac{k}{1-\beta_p}+1)\rceil+ \lceil\frac{2k}{1-\beta_p}\rceil.$$
\end{Lemma}
\proof
1) $I_{(k_0\geq k_1)}$ can be represented as $1$ bit.

2) Without loss of generality, we assume $k_0\leq k_1$. According to our construction,
$$ \log_2 {\nchoosek{k_0+k_1-1}{ k_0-1}}< \frac{k}{1-\beta_p} \textrm{ for } k_0>1,$$
and
$$ \log_2 {\nchoosek{k_1}{1}} < \frac{k}{1-\beta_p} \textrm{ for } k_0=0 \textrm{ or } k_0=1.$$

Then
\begin{eqnarray*}
k_0-1&\leq &\log_2 {\nchoosek{2k_0-1}{ k_0-1}}\\
&\leq& \log_2 {\nchoosek{k_0+k_1-1}{k_0-1}}\\
&<& \frac{k}{1-\beta_p}.
\end{eqnarray*}

So $\min(k_0, k_1)$ can be represented as
$\lceil\log_2 (\frac{k}{1-\beta_p} +1 )\rceil$ bits.

3) Let us consider the number of permutations of $x$, denoted by $N(x)$.
If $k_0>1$, then
\begin{eqnarray*}
\log_2 N(x)&=&\log_2 {\nchoosek{k_0+k_1}{ k_0}}\\
&\leq &\log_2  {\nchoosek{k_0+k_1-1}{ k_0-1}} +\log_2\frac{k_0+k_1}{k_0}\\
&\leq &\frac{k}{1-\beta_p} +\log_2\frac{k_0+k_1}{k_0}.
\end{eqnarray*}

If $k_0=1$, then
$$\log_2 N(x) \leq \log_2  {\nchoosek{k_1 }{1}} +\log_2 \frac{k_1+1}{k_1}.$$

If $k_0=0$, then
$$\log_2 N(x)=0.$$

Based on the analysis above, we can get $$\log_2 N(x)\leq \frac{2k}{1-\beta_p}.$$

Hence, $r(x)$ can be represented as $\lceil\frac{2k}{1-\beta_p}\rceil$ bits.

This completes the proof.
\hfill\QED

Let $\mathbf{1}^a$ denote the all-one vector of length $a$, then we get the following result.
\begin{Theorem}
 Construction \ref{const:2} generates a random sequence of length $m$ that is $\epsilon$-close to $U_m$ if
 $P_{\mathcal{R}}(\mathbf{1}^a)\leq 2^{-k}, P_{\mathcal{R}}(\mathbf{0}^a)\leq 2^{-k}$ for $a=2^{\lfloor\frac{k}{1-\beta_p}\rfloor}$.
\end{Theorem}

\proof Since the mapping in the second step is injective, it will not affect the min-entropy; we only need to prove that the input sequence has min-entropy $k$, i.e.,
$$\log_2 \frac{1}{P_{\mathcal{R}}(x)}\geq k, \forall x\in S_p,$$
where $S_p$ is the set consisting of all the possible input sequences.

It is not hard to see that if $\min(k_0,k_1)\geq 1$,
$$P_{\mathcal{M}}(x)\leq \frac{1}{{\nchoosek{k_0+k_1}{ k_0}}},$$
which leads to
$$\log_2 \frac{1}{P_{\mathcal{M}}(x)}\geq \frac{k}{1-\beta_p}.$$

Furthermore, based on the definition of $d_p(\mathcal{R},\mathcal{M})$, we can get if $\min(k_0,k_1)\geq 1$,
$$\log_2 \frac{1}{P_{\mathcal{R}}(x)}\geq k.$$

If $\min(k_0,k_1)=0$, according to the condition in the lemma, we can also have the same result.

Since $k=m(1+\alpha)$ with $\alpha>0$, according to Lemma \ref{lemma_seededextractor},
we can construct a seeded extractor that applies to the sequence $Z$ and generates a binary sequence $\epsilon$-close to the uniform distribution $U_m$.

This completes the proof.
\hfill\QED

Actually, the idea above can be easily generalized if $\mathcal{M}$ is a Markov chain that best approximates the real source $\mathcal{R}$.
The idea follows the main lemma in \cite{Zhou_Markov} that shows how to generate random bits with optimal efficiency from an arbitrary Markov chain.

\subsection{Efficiency Analysis}

For the efficiency of the construction, we can get the same bounds as Construction \ref{const:1}.

\begin{Theorem}
Given an arbitrary source $\mathcal{R}$ such that $$\min_{\mathcal{M}\in \mathcal{G}_{b.c.}}d (\mathcal{R},\mathcal{M}) \leq \beta,$$
if there exists a model $\mathcal{M}\in \mathcal{G}_{b.c.}$ with probability $p\leq \frac{1}{2}$ of being $1$ or $0$ and
$$p>\sqrt{d(\mathcal{R},\mathcal{M})\log_2\frac{1}{p}\frac{\ln 2}{2}},$$ then the efficiency of Construction \ref{const:2} is
$$1-\beta \leq \eta\leq 1.$$
\end{Theorem}

\proof Let $N_{k_0, k_1}$ denote the number of input sequences with $k_0$ zeros and $k_1$ ones in $S_p$, and let $p_{k_0,k_1}$ be the probability based on $\mathcal{R}$ of generating such a sequence. Let us define
$$A=\{(k_0,k_1)| N_{k_0,k_1}>0\},$$
then we can get
$$
H_{\mathcal{R}}(X_m)\leq H(\{p_{k_0,k_1}|(k_0,k_1)\in A\})$$
$$+\sum_{(k_0,k_1)\in A}p_{k_0,k_1}\log_2 N_{k_0,k_1}.
$$

According to the proof in the above theorem, $\min(k_0,k_1)\leq \frac{k}{1-\beta_p}+1$. So there are totally at most $2(\frac{k}{1-\beta_p}+1)$ available pairs of $(k_0, k_1)$. Hence
$$H(\{p_{k_0,k_1}|(k_0,k_1)\in A\})\leq \log_2(2+(\frac{k}{1-\beta_p}+1))=o(k).$$

Now, we write $n=k_0+k_1$. According to our method, if $\min(k_0,k_1)\geq 1$,
$$\nchoosek{k_0+k_1}{\min(k_0,k_1)}\geq 2^{\frac{k}{1-\beta_p}},$$
$$\nchoosek{k_0+k_1-1}{\min(k_0,k_1)-1}<2^{\frac{k}{1-\beta_p}}.$$

Hence, given $n$, we get an upper bound for $\min(k_0,k_1)$, which is
\begin{equation}\label{var_equ_lemma1_1}
t_n=\max\{i\in\{0,1,...,n\}| {\nchoosek{n-1}{ i-1}}<2^{\frac{k}{1-\beta_p}}\}.
\end{equation}

Note that if $\nchoosek{n-1}{\frac{n}{2}-1}\geq 2^{\frac{k}{1-\beta_p}}$, then $t_n$ is a nondecreasing
function of $n$. Using the Stirling bounds on factorials yields
$$ \lim_{n\rightarrow \infty }\frac{1}{n}\log_2 {\nchoosek{n}{ \rho n}}=H(\rho),$$
where $H$ is the binary entropy function. Hence, following (\ref{var_equ_lemma1_1}), we can get
\begin{equation}\label{var_equ_lemma1_3} \lim_{n\rightarrow \infty} H(\frac{t_n}{n})=\lim_{n\rightarrow\infty}\frac{k}{(1-\beta_p) n}.\end{equation}

Let $P_n$ denote the probability of having an input sequence of length at least $n$ based on the distribution of $\mathcal{R}$. In this case,
$P_n$ is a nonincreasing function of $n$. Let $Q_n$ denote the probability of having an input sequence of length at least $n$ based on the distribution of $\mathcal{M}\in \mathcal{G}_{b.c.}$
whose probability is $p\leq \frac{1}{2}$. Since for all binary sequence $x\in \{0,1\}^n$,
$$\log_2 \frac{1}{P_\mathcal{M}(x)}\leq n\log_2\frac{1}{p},$$
we can get
$$\log_2\frac{P_{\mathcal{R}(x)}}{P_\mathcal{M}(x)}\leq d n\log_2 \frac{1}{p},$$
where $d=d_p(\mathcal{R},\mathcal{M})$.

Since
$P_n=\sum_{x\in S} P_{\mathcal{R}}(x)$ and $Q_n=\sum_{x\in S}
P_{\mathcal{M}}(x)$ for some $S\subset \{0,1\}^n$, it is not hard to prove that
\begin{equation}\label{equ_lemma1_13}\log_2\frac{P_n}{Q_n}\leq d n \log_2\frac{1}{p}.\end{equation}

According to Hoeffding's inequality, we can get
\begin{eqnarray*}
Q_n&\leq &2P[k_1\leq t_n]\\
&\leq& 2P[\frac{k_1}{n}-p \leq \frac{t_n}{n}-p] \\
&\leq& 2e^{-2n(p-\frac{t_n}{n})^2}.
\end{eqnarray*}

Hence
\begin{equation}\label{equ_lemma1_14}P_n \leq 2^{- d n\log_2 p} Q_n \leq 2 e^{-\log_2 p \ln 2 \cdot d n  -2 n (p-\frac{t_n}{n})^2}.\end{equation}

From this inequality, we see that $P_n\rightarrow 0$ as $n\rightarrow 0$ if
\begin{equation}\label{equ_lemma1_15}-d \log_2 p \ln 2 -2 (p-\frac{t_n}{n})^2 <0.\end{equation}

Based on (\ref{var_equ_lemma1_3}) and (\ref{equ_lemma1_15}), we can get that $P_n\rightarrow 0$ as $n\rightarrow 0$ if
$$\frac{n}{k}\geq \frac{1}{(1-\beta_p)H(p-\sqrt{d\log_2\frac{1}{p}\frac{\ln 2}{2}})}.$$

Now, let $a=\frac{1+\epsilon}{(1-\beta_p)H(p-\sqrt{d\log_2\frac{1}{p}\frac{\ln 2}{2}})}$ with $\epsilon>0$, we can write
\begin{eqnarray*}
H_{\mathcal{R}}(X_m) &\leq &o(k)+ \sum_{k_0,k_1: k_0+k_1\geq a k} p_{k_0,k_1}\log_2 N_{k_0,k_1}\\
&&+ \sum_{k_0,k_1: k_0+k_1<a k} p_{k_0,k_1}\log_2 N_{k_0,k_1}.
\end{eqnarray*}

According to our analysis, if $k_0+k_1\geq a k$, as $k\rightarrow \infty$, $$P_n=\sum_{k_0,k_1:k_0+k_1\geq a k} p_{k_0,k_1}\rightarrow 0$$ and
$\log_2 N_{k_0,k_1}\leq 2\frac{k}{1-\beta_p}$.  If $k_0+k_1\leq a k$, then $$\log_2 N_{k_0,k_1}\leq \frac{k}{1-\beta_p}+\log_2 \frac{k_0+k_1}{\min (k_0,k_1)}\leq \frac{k}{1-\beta_p} +o(k).$$ As a result, we can get
\begin{eqnarray*}
H_{\mathcal{R}}(X_m)& \leq &o(k)+ o(1)\frac{2k}{1-\beta_p} + (\frac{k}{1-\beta_p}+o(k))\\
&\leq& \frac{k}{1-\beta_p}+o(k).
\end{eqnarray*}

So
$$\lim_{k\rightarrow
\infty}\frac{k}{H_{\mathcal{R}}(X_m)}\geq 1-\beta.$$

Furthermore, based on the fact that $\lim_{m\rightarrow\infty}\frac{k}{m}=1$, we can get $\eta\geq 1-\beta$.
It is known that $\eta\leq 1$, so it concludes the theorem.
\hfill\QED

Similar to Construction \ref{const:1}, this construction is also asymptotically optimal in the sense
that we cannot find a variable-length extractor with efficiency definitely larger than $1-\beta$, as shown in Theorem \ref{var_theorem_lowerbound1}.

\begin{Corollary} Given an arbitrary source $\mathcal{R}$ such that $$\min_{\mathcal{M}\in \mathcal{G}_{b.c.}} d(\mathcal{R},\mathcal{M}) \leq \beta,$$ then as $\beta\rightarrow 0$, the efficiency of Construction \ref{const:2} is
$$\eta\rightarrow 1.$$
\end{Corollary}

It is easy to see that as $\beta\rightarrow 0$, Construction \ref{const:2} reaches the Shannon's limit on efficiency. If
$\mathcal{R}$ is a stationary ergodic process, we can also get the following corollary.

\begin{Corollary}
Given an arbitrary stationary ergodic source $\mathcal{R}$ such that $$\min_{\mathcal{M}\in \mathcal{G}_{b.c.}} d(\mathcal{R},\mathcal{M}) \leq \beta,$$
if there exists a model $\mathcal{M}\in \mathcal{G}_{b.c.}$ with probability $p\leq \frac{1}{2}$ of being $1$ or $0$ and
$$p>\sqrt{d(\mathcal{R},\mathcal{M})\log_2\frac{1}{p}\frac{\ln 2}{2}},$$ then for the expected input length of Construction \ref{const:2}, we have
$$\frac{1}{h(\mathcal{R})}\leq \lim_{m\rightarrow\infty} \frac{E[|X_m|]}{m}\leq \frac{1}{(1-\beta)h(\mathcal{R})},$$
where $h(\mathcal{R})$ is the entropy rate of $\mathcal{R}$.
\end{Corollary}

\section{Construction III: Approximately Stationary Ergodic Processes}
\label{var_section_tech4}

In this section, we consider imperfect sources that are approximately stationary and ergodic.
Here, we let  $\mathcal{R}$ be an arbitrary stochastic process such that $d(\mathcal{R},\mathcal{M})\leq \beta$ for a stationary ergodic process $\mathcal{M}$.
For these sources, universal data compression can be used to `purify' input sequences, i.e., shortening their lengths while maintaining their entropies.
In \cite{Visweswariah98}, Visweswariah, Kulkarni and Verd\'{u} showed that optimal variable-length source codes asymptotically achieve optimal variable-length random bits generation in the sense of normalized divergence. Although
their work only focused on ideal stationary ergodic processes and generates `weaker' random bits, it motivates us to combine
universal compression with fixed-length extractors for efficiently generating random bits from noisy stochastic processes.
In this section, we will first introduce Lempel-Ziv code and then present its application in constructing variable-length extractors.

\subsection{Construction}

Lempel-Ziv code is a universal data compression scheme introduced by Ziv and Lempel \cite{Ziv78}, which is simple to implement and can achieve the asymptotically optimal rate
for stationary ergodic sources. The idea of Lempel-Ziv code is to parse the source sequence into strings that have not appeared so far, as demonstrated by the following example.

\begin{Example} Assume the input is $010111001110000...$, then we parse it as strings
$$0,1,01,11,00,111,000,...$$
where each string is the shortest string that never appear before. That means all its prefixes have occurred earlier.

Let $c(n)$ be the number of strings obtained by parsing a sequence of length $n$. For each string, we
describe its location with $\log c(n)$ bits. Given a string of length $l$, it can described by (1)
the location of its prefix of length $l-1$, and (2) its last bit. Hence, the code for the above sequence is
{ $$(000,0), (000,1), (001,1), (010,1), (001,0), (100,1), (101,0),...$$}
where the first number in each pair indicates the prefix location and the second number is
the last bit of the string.
\end{Example}
\vspace{-0.25cm}
\hfill$\Box$

Typically, Lempel-Ziv is applied to an input sequence of fixed length. Here, we are interested in
Lempel-Ziv code with fixed output length and variable input length. As a result, we can apply
a single fixed-length extractor to the output of Lempel-Ziv code for extracting randomness.
In our algorithm, we read raw bits one by one from an imperfect source until the length of the output
of a Lempel-Ziv code reaches a certain length.
In another word,
the number of strings after parsing is a predetermined number $c$. For example, if
the source is $1011010100010...$ and $c=4$, then after reading $6$ bits, we can parse them into
$1, 0, 11, 01$. Now, we get an output sequence
$(000,1), (000,0), (001, 1), (010, 1)$, which can be used as the input of a fixed-length extractor.
We call this Lempel-Ziv code as a variable-length Lempel-Ziv code.

Let $Z$ be a random sequence obtained based on variable-length Lempel-Ziv code such that its length is
$$|Z|=(\log c+1) c,$$
for a predetermined $c$. Then $Z$ is very close to truly random bits in the term of min-entropy if the source $\mathcal{R}$ is stationary ergodic.
As a result, we have the following construction for variable-length extractors.

\begin{Construction}\label{const:3}
Assume the real source is $\mathcal{R}$ and there exists a stationary ergodic process $\mathcal{M}$ such that $d(\mathcal{R},M)\leq \beta$.
Then we extract $m$ almost random bits from $\mathcal{R}$ based on the following procedure.
\begin{enumerate}
\item Read input bits one by one based on the variable-length Lempel-Ziv code until we get an output sequence $Z$ whose length reaches $$n=\frac{k}{1-\beta_p}(1+\varepsilon),$$
    where $\varepsilon>0$ is a small constant indicating the performance gap between the case of finite-length and that of infinite-length
    for Lempel-Ziv code; as $m\rightarrow\infty$, we have $\varepsilon\rightarrow 0$. Similar as above,
    $\beta_p=\beta+\epsilon_p$ with $\epsilon_p>0$ and $k=m(1+\alpha)$ with $\alpha>0$. The small constant $\epsilon_p$ has value depending on the input set $S_p$; as
  $m\rightarrow\infty$, $\epsilon_p\rightarrow 0$. The constant $\alpha$ can be arbitrarily small.
    Then we get a random sequence $Z$ of length $n$ and with min-entropy $k$.
\item Applying a $(k,\epsilon)$ extractor to $Z$ yields a random sequence of length $m$ that is $\epsilon$-close to $U_m$.\hfill$\Box$
\end{enumerate}
\end{Construction}

We show that the min-entropy of $Z$ is at least $k$ as $m\rightarrow\infty$.  If $m$ is not very large, by adjusting the parameter $\varepsilon$,
we can make the min-entropy of $Z$ be at least $k$.
So we can continue to apply
an efficient fixed-length extractor to `purify' the resulting sequence. Finally, we can get $m$
random bits that satisfy our requirements on quality in the sense of statistical distance.

\begin{Theorem}\label{theorem_3_1}
When $m\rightarrow\infty$, Construction \ref{const:3} generates a random sequence of length $m$ that is $\epsilon$-close to $U_m$.
\end{Theorem}

\proof Let $x$ be an input sequence. According to theorem 12.10.1 in \cite{Cover2006}, for the stationary ergodic process $\mathcal{M}$,
we can get
$$\frac{1}{|x|}\log_2 \frac{1}{ P_{\mathcal{M}}(x)}\geq \frac{c}{|x|}\log_2 c -\frac{c}{|x|} H(U,V),$$
where  $$\frac{c}{|x|}H(U,V)\rightarrow 0 \textrm{ as } |x|\rightarrow 0.$$

As a result, if $k=\Theta(n)$,
\begin{eqnarray*}
\lim_{k\rightarrow\infty}\frac{1}{k}\log_2 \frac{1}{ P_{\mathcal{R}}(x)}&\geq& \lim_{k\rightarrow\infty}(1-\beta_p)\frac{1}{k}\log_2\frac{1}{P_\mathcal{M}(x)}\\
& \geq &\lim_{k\rightarrow\infty} \frac{(1-\beta_p)c\log_2 c}{k} \\
&=&  \lim_{k\rightarrow \infty}\frac{(1-\beta_p)n}{k}\\
&=& \lim_{k\rightarrow\infty} 1+\varepsilon\\
&=& 1.
\end{eqnarray*}

Finally, we can get that
$$\lim_{k\rightarrow\infty} \frac{H_{\min}(Z)}{k}=\lim_{k\rightarrow\infty} \frac{H_{\min}(X_m)}{k}\geq 1.$$
This implies that as $m\rightarrow \infty$, i.e., $k\rightarrow \infty$, the min-entropy of $Z$ is at least $k$.

Since $k=m(1+\alpha)$ for an $\alpha>0$,  we can continue to apply a $(k,\epsilon)$ extractor to extract $m$ almost-random bits from $Z$.
\hfill\QED

\subsection{Efficiency Analysis}

Now, we study the efficiency of the construction based on variable-length Lempel-Ziv codes.

\begin{Theorem}
Given a real source $\mathcal{R}$ such that there exists a stationary ergodic process $\mathcal{M}$ with $d(\mathcal{R},\mathcal{M})\leq \beta$, then the efficiency of Construction \ref{const:3} is \vspace{-0.05cm}
$$1-\beta\leq \eta\leq 1.$$
\end{Theorem}

\proof Similar as above, we only need to prove that $\eta\geq 1-\beta$.

Since there are at most $n=2^{c(\log_2c+1)}$ distinct input sequences,
their entropy
$$H_{\mathcal{R}}(X_m)\leq c(\log_2 c+1)=n.$$

According to the proof in Theorem \ref{theorem_3_1}, we have that the random sequence $Z$ has min-entropy at least $k$, and it satisfies
$$\lim_{m\rightarrow\infty}\frac{n}{k}=\frac{1}{1-\beta}.$$

Based on the construction of seeded extractors, we can also get
$$\lim_{m\rightarrow\infty}\frac{m}{k}=1.$$

As a result,
$$ \eta= \lim_{m\rightarrow\infty}\frac{m}{H_{\mathcal{R}}(X_m)} \geq 1-\beta.$$
This completes the proof.
\hfill\QED

Although Construction \ref{const:3} has the same efficiency as the other constructions, when $m$ is not large,
it is  less efficient than the other constructions because the Lempel-Ziv code does not always have the best performance when the input sequence is not long.
But its advantage is that it can manage more general sources without accurate estimations. In the above theorem, the gap $\beta$ represents how far the source $\mathcal{R}$ is from being stationary ergodic. In general, the efficiency loss introduced by the uncertainty of sources
is a part that cannot be avoid.

\begin{Corollary} Given a real source $\mathcal{R}$ such that there exists a stationary ergodic model $\mathcal{M}$ with $d(\mathcal{R},\mathcal{M})\leq \beta$, then as $\beta\rightarrow 0$, the efficiency of Construction \ref{const:3} is
$$\eta\rightarrow 1.$$
\end{Corollary}

It shows that as $\beta\rightarrow 0$, Construction \ref{const:3} reaches the Shannon's limit on efficiency.

\begin{Corollary} Given a stationary ergodic source $\mathcal{R}$ (assume we do not know that it is stationary ergodic), for the expected input length of Construction \ref{const:3}, we have
$$\frac{1}{h(\mathcal{R})}\leq \lim_{m\rightarrow\infty} \frac{E[|X_m|]}{m}\leq \frac{1}{(1-\beta)h(\mathcal{R})},$$
where $h(\mathcal{R})$ is the entropy rate of $\mathcal{R}$.
\end{Corollary}

\section{Seedless Constructions}
\label{sec_randomnessextraction}

To simulate seeded constructions of variable-length extractors in randomized applications, we have to enumerate all possible assignments of the seed, hence,
the computational complexity will be increased significantly. In real applications, we prefer seedless constructions rather than seeded constructions. It motivates us to study the seedless constructions of variable-length extractors in this section.

\subsection{An Independent Source}

Let us first consider a simple independent source described in the introduction.  This type of sources have been widely studied in seedless constructions of fixed-length extractors.

\begin{Example} Let $x_1x_2...\in\{0,1\}^*$ be an independent sequence generated from a source $\mathcal{R}$ such that
$$P[x_i=1]\in [0.9,0.91] \quad \forall i\geq i.$$
\end{Example}
\vspace{-0.4cm} \hfill$\Box$
\vspace{0.2cm}

We see that the existing methods for generating random bits from ideal sources (like biased coins or Markov chains) cannot be applied here, since
the probability of each bit is slightly unpredictable. Some seedless extractors have been developed for extracting randomness from such sources.
In particular, there exists seedless extractors which are able to extract as many as $H_{\min}(X)$ random bits from a independent random sequence $X$ asymptotically. In order to extract $m$ random bits in the above example, it needs to read $\frac{m}{\log_2\frac{1}{0.91}}$ input bits as
$m\rightarrow\infty$. In this case, the entropy of the input sequence is in $$[H(0.9)\frac{m}{\log_2\frac{1}{0.91}}, H(0.91)\frac{m}{\log_2\frac{1}{0.91}}].$$ From which, we can get the efficiency of an optimal fixed-length  extractor, which is
$$\eta_{fixed}\in [0.2901, 0.3117],$$
i.e., about only $0.3$ of the input entropy is used for generating random bits, which is far from optimal

In the above example, we let $\mathcal{M}$ be a biased coin model with probability $p=0.9072$. In this case,
$$\beta\leq d(\mathcal{R},\mathcal{M})=0.0315.$$
According to the constructions in the previous sections, there exists seeded variable-length extractors such that
their efficiencies are
$$\eta_{variable}\in [1-\beta, 1]\subseteq [0.9685,1],$$
which are near Shannon's limit.

Based on the fact that the source is independent, we can eliminate the requirement of truly random bits as the seed, hence, we have seedless
variable-length extractors.
To construct a seedless variable-length extractor, we first apply a seedless fixed-length extractor $E_1$ (which may not be very efficient)
to extract a random sequence of length $d$ from input bits.
Using this random sequence as the seed, we continue to apply a seeded variable-length extractors $E_2$ to
extract $m$ almost-random bits from extra input bits. So seedless variable-length extractors can be constructed as cascades of seedless fixed-length extractors and seeded variable-length extractors. Since the input length of $E_1$ is much shorter (it is ignorable) than the input length
of $E_2$, the efficiency of the resulting seedless extractor, i.e., $E=E_2\bigotimes E_1$, is dominated by the efficiency of $E_2$.
So the efficiency of the seedless extractor $E$ is in $[0.9685,1]$,
which is very close to the optimality.

This example demonstrates a simple construction of seedless variable-length extractors for independent sources, and it shows
the significant performance gain of variable-length extractors compared to fixed-length extractors.

\subsection{Generalized Sources}

Here we consider a generalization of independent processes. Given a system, we use $\lambda_i$ denote the complete system status at time $i$.
For example, in a system that generates thermal noise, the system status can include the value of the noise signal, the temperature, the environmental effects, etc. Usually, the evolution of such a system has a Markov property, namely,
$$P[\lambda_{i+1},\lambda_{i+2},...|\lambda_i, \lambda_{i-1},..., \lambda_1]=P[\lambda_{i+1},\lambda_{i+2},...|\lambda_i],$$
for all $i\geq 1$. Let $X=x_1x_2...\in \{0,1\}^n$ be the binary sequence generated from this system, then for any $1< k<n-1$,
\begin{equation}\label{equ_seedless_1}P[X_1^{k-1},X_{k+1}^{n}|\lambda_k]=P[X_1^{k-1}|\lambda_k]P[X_{k+1}^{n}|\lambda_k],\end{equation}
where $X_a^b=x_ax_{a+1}...x_{b}$. In some sense, the source $X$ that we consider is a hidden Markov process, but the number of hidden states can be infinite ($\lambda_i$ can be discrete or continuous).

\begin{Example} One example of the above sources is the one studied in \cite{Kamp11}, called a space $s$ source.
A space $s$ source is basically a source generated by a width $2^s$ branching program. At each step, the state of the process
generating the source is in one of $2^s$ states, and the bit generated is a function of the current state. Unlike perfect Markov chains,
the transition probabilities can be different at each step. In this example, the system status $\lambda_i$ is the content of space $s$ at time $i$, that is, one of the $2^s$ states, and $x_i\in \{0,1\}$ is a function of $\lambda_i$.
\end{Example}

Space $s$ sources are very general in that most other classes of sources that have been considered
previously can be computed with a small amount of space \cite{Kamp11}. The model that we consider, as described by (\ref{equ_seedless_1}),
is a natural generalization of space $s$ sources. This model has a very nice feature: from such a source, we can get a group of sequences conditionally independent of each other. Namely,
given system statues at some time points
$$[\lambda^{(1)}, \lambda^{(2)}, ..., \lambda^{(\gamma)}]=[\lambda_{a}, \lambda_{2a}, ..., \lambda_{\gamma a}],$$
the subsequences
$$[X^{(1)},X^{(2)},...,X^{(\gamma)}, X^{(\gamma+1)}]$$
$$=[X_1^{a-1}, X_{a+1}^{2a-1}, ..., X_{(\gamma-1)a+1}^{\gamma a-1}, X_{\gamma a+1}^\infty]$$
are conditionally independent of each other. Based on this condition, we have the following seedless construction of
variable-length extractors.

\begin{Construction}\label{const:4}
Given a source $\mathcal{R}$ described by (\ref{equ_seedless_1}), we can construct a seedless variable-length extractor $E$ in the following way:
\begin{enumerate}
  \item Suppose that $$H_{\min}(X^{(i)}|\lambda^{(i)}, \lambda^{(i+1)})\geq k_{d}, \forall 0\leq i\leq \gamma.$$
  We construct a $\gamma$-source extractor \cite{Rao2007} $E_1:[\{0,1\}^{a-1}]^\gamma \rightarrow \{0,1\}^d$ such that if each source has
  min-entropy $k_d$, it can extract $d$ almost-random bits which are $\epsilon_1$-close to the uniform distribution on $\{0,1\}^d$.
  \item We construct a seeded variable-length extractor $E_2: S_p\times \{0,1\}^d\rightarrow\{0,1\}^m$ such that
  with condition on $\lambda^{(\gamma)}$,
  it can extract $m$ almost-random bits from $X^{(\gamma+1)}$ and these $m$ almost-random bits are $\epsilon_2$-close to the uniform distribution on
  $\{0,1\}^m$ if the seed is truly random.
  \item The seedless variable-length extractor $E$ is a cascade of $E_1$ and $E_2$:
  Let $$D=E_1(X^{(1)}, X^{(2)},..., X^{(\gamma)}),$$
  then we apply $D$ as the seed of $E_2$ to generate $m$ almost-random bits from $X^{(\gamma+1)}$; that is,
  $$E(X)=E_2(X^{(\gamma+1)}, E_1(X^{(1)}, X^{(2)},..., X^{(\gamma)})).$$
  \end{enumerate}
\end{Construction}

For this construction, we have the following theorems.

\begin{Theorem} In Construction \ref{const:4}, the $m$ almost-random bits generated by the seedless variable-length extractor $E$ are $(\epsilon_1+\epsilon_2)$-close to the uniform
distribution on $\{0,1\}^m$.
\end{Theorem}

\proof According to the construction, we can let the parameter $a=|X^{(i)}|+1$ with $1\leq i\leq \gamma$ be large enough, so given $\lambda^{(1)}, \lambda^{(2)}, ..., \lambda^{(\gamma)}$,
$$\|D-U_d\|\leq \epsilon_1.$$

Let $X_m$ be the input sequence of $E_2$ that read from $X^{(\gamma+1)}$, then given  $\lambda^{(\gamma)}$,
we have
$$\|E_2(X_m, U_d)-U_m\|\leq \epsilon_2.$$

From the two inequalities above, given $\lambda^{(1)}, \lambda^{(2)}, ..., \lambda^{(\gamma)}$,
we have
$$\|E_2(X_m, D)-U_m\|\leq \epsilon_1+\epsilon_2.$$

Since it is true for any assignments of  $\lambda^{(1)}, \lambda^{(2)}, ..., \lambda^{(\gamma)}$, we can get
\begin{eqnarray*}
  &&  \|E_2(X_m, D)-U_m\|\\
   &=& \sum_{\lambda^{(1)}, \lambda^{(2)}, ..., \lambda^{(\gamma)}}P[\lambda^{(1)}, \lambda^{(2)}, ..., \lambda^{(\gamma)}] (\epsilon_1+\epsilon_2)\\
   &\leq & \epsilon_1+\epsilon_2.
\end{eqnarray*}

Hence, the $m$ almost-random bits extracted by $E$ is also $(\epsilon_1+\epsilon_2)$-close to $U_m$.
\hfill\QED

In the following theorem, we show that the seedless variable-length extractor $E$ has the efficiency as
the seeded variable-length extractor $E_2$.

\begin{Theorem} In Construction \ref{const:4}, suppose that
$$H_{\min}(X^{(i)}|\lambda^{(i)}, \lambda^{(i+1)})=\Theta(|X^{(i)}|), \forall 0\leq i\leq \gamma.$$
Let $\eta_{E}$ denote the efficiency of the resulting seedless variable-length extractor $E$, and
let $\eta_{E_2}$ denote the efficiency of the $E_2$, then
$$\eta_E=\eta_{E_2}.$$
\end{Theorem}

\proof According to the construction of $E_1$, we can get that
$$d=\Theta(a),$$
where $a=|X^{(i)}|+1$ for $1\leq i\leq \gamma$.

If $\epsilon_2$ is a constant, then
$$d=O(\log m)=o(m).$$

As a result, $$\lim_{m\rightarrow\infty}\frac{a\gamma}{m}=0.$$
 Let $H$ denote the entropy of the input sequence of $E_2$,
then
$\eta_{E_2}=\lim_{m\rightarrow\infty}\frac{m}{H}$, and
$$\lim_{m\rightarrow\infty}\frac{m}{a\gamma + H} \leq \eta_E \leq \lim_{m\rightarrow\infty}\frac{m}{H}.$$

Hence, $\eta_E=\eta_{E_2}$.
\hfill\QED

The theorem above shows that the efficiency of seedless variable-length extractors can be very close to optimality. For many sources, such as biased coins with noise, or Markov chains with noise, the existing algorithms for ideal sources (e.g., perfect biased coins or perfect Markov chains)
cannot generate high-quality random bits from them. At the same time, the traditional approaches of fixed-length extractors are not very efficient. The gap between their efficiency and the optimality is determined by the bias of the source. Seedless variable-length extractors take the advantages of both, as a result, they can approach the information-theoretic upper bound on efficiency while being capable to combat noise in the sources.

\section{Conclusion and Discussion}

In this paper, we introduced the concept of the variable-length extractors, namely, those extractors with variable input length and
fixed output length. Variable-length extractors are generalizations of the existing algorithms for ideal sources to manage general stochastic processes. They are also improvements of traditional fixed-length extractors to fill the gap between min-entropy and entropy of the source
on efficiency. The key idea of constructing variable-length extractors is to approximate the source using a simple model, which is a known process,
a biased coin, or a stationary ergodic process. Depending on the model selected, we proposed and analyzed three seeded constructions
of variable-length extractors. Their efficiency is lower bounded by $1-\beta$ and upper bounded by $1$ (optimality), where
$\beta (0\leq \beta \leq 1)$ indicate the uncertainty of the real source. We also show that our constructions are asymptotically optimal, in the sense that one cannot find a construction whose efficiency is always strictly larger than $1-\beta$.
 In addition, we demonstrated how to construct
seedless variable-length extractors by cascading seeded variable-length extractors with seedless fixed-length extractors. They can work for many (but not all) natural sources such as those based on noise signals.

There are certain connections between variable-length extractors and a whole family of variable-to-fixed length source codes \cite{Lawrence1977,Tjalkens1992,Merhav1992,Savari1997,Tjalkens1987,Tunstall1967,Visweswariah2001}. With a variable-to-fixed length code, an infinite sequence is parsed into variable-length phases, chosen from some finite set $\mathcal{D}$ of phases. Each phase is then
coded into a binary sequence of fixed length $m$. The set $\mathcal{D}$ of phases is complete, i.e., every infinite sequence has a prefix in $\mathcal{D}$. The key of constructing a good variable-to-fixed length source code is to find the best set $\mathcal{D}$ that consists of at last $2^m$ prefix-free phases and maximizes the expected phase length. As comparison, the key of constructing a variable-length extractor is to find
the best input set $S_p$ that consists of sequences with probability at most $2^{-k}$ for each and minimizes the expected sequence length.
Although their goals are different, some common ideas can be used to construct both the phase set $\mathcal{D}$ and the input set $S_p$.
For example, in \cite{Visweswariah2001}, Visweswariah et al.  defined  the phase set $\mathcal{D}$ by $x^*\in \mathcal{D}$ if $P(x^*)\leq c$ and no prefix of $x^*$ satisfies this property. The same idea is applied in our construction I.  In \cite{Lawrence1977,Tjalkens1992},
the phase set $\mathcal{D}$ is determined by the number of ones and zeros in the phase, so is our construction II.  In some sense,
an optimal variable-to-fixed length code can result in a fixed-length binary sequence whose min-entropy is close to its length. However,
variable-to-fixed length source codes do not always work well in constructing variable-length extractors, because
(1) the designing criteria are different and they may degrade the performance; (2) variable-to-fixed length source codes take both encoding and decoding in consideration, hence, they are more complex in computation than what we require (decoding is not necessary) for constructing variable-length extractors; and (3) the sources that we considered for variable-length extractors are unpredictable, which are more general than the ones considered in variable-to-fixed length source codes.

\end{document}